\documentclass[conference]{IEEEtran}

\usepackage{gianluca}

\usepackage{pdfpages}
\usepackage{enumitem}
\setlist[itemize]{leftmargin=*}

\usepackage[colorinlistoftodos]{todonotes}

\usepackage{ulem}
\usepackage{endnotes}

\usepackage[noadjust]{cite}

\begin{document}


\DeclareRobustCommand*{\IEEEauthorrefmark}[1]{%
  \raisebox{0pt}[0pt][0pt]{\textsuperscript{\footnotesize #1}}%
}

\title{DeepFloat: Resource-Efficient Dynamic Management of Vehicular Floating Content}

\author{
    \IEEEauthorblockN{Gaetano Manzo\IEEEauthorrefmark{1,2}, Sebastian Ot\'{a}lora\IEEEauthorrefmark{1}, Marco Ajmone Marsan\IEEEauthorrefmark{3}, Torsten Braun\IEEEauthorrefmark{2}, Hung Nguyen \IEEEauthorrefmark{4}, Gianluca Rizzo\IEEEauthorrefmark{1}}
    \IEEEauthorblockA{\IEEEauthorrefmark{1}University of Applied Science Western Switzerland $-$
    \{name.surname\}@hevs.ch}
    \IEEEauthorblockA{\IEEEauthorrefmark{2}University of Bern, Switzerland
    $-$ braun@inf.unibe.ch}
     \IEEEauthorblockA{\IEEEauthorrefmark{3} IMDEA Networks Institute, Spain \& Politecnico di Torino, Italy
    $-$ ajmone@polito.it}
     \IEEEauthorblockA{\IEEEauthorrefmark{4} University of Adelaide, Australia
    $-$ hung.nguyen@adelaide.edu.au}
    
}

\maketitle
\begin{abstract}

Opportunistic communications are expected to play a crucial role in enabling context-aware vehicular services.
A widely investigated opportunistic communication paradigm for storing a piece of content probabilistically in a geographical area is Floating Content (FC). 
A key issue in the practical deployment of FC is how to tune content replication and caching in a way which achieves a target performance (in terms of the mean fraction of users possessing the content in a given region of space) while minimizing the use of bandwidth and host memory.
Fully distributed, distance-based approaches prove highly inefficient, and may not meet the performance target, while centralized, model-based approaches do not perform well in realistic, inhomogeneous settings. 

In this work, we present a data-driven centralized approach to resource-efficient, QoS-aware dynamic management of FC. 
We propose a Deep Learning strategy, which employs a Convolutional Neural Network (CNN) to capture the relationships between patterns of users mobility, of content diffusion and replication, and FC performance in terms of resource utilization and of content availability within a given area. 
Numerical evaluations show the effectiveness of our approach in deriving strategies which efficiently modulate the FC operation in space and  effectively adapt to mobility pattern changes over time. 

\end{abstract}

\section{Introduction}
Vehicular communications are considered one of the key vertical application domains for 5G \cite{5Gverticals}, and a version of the IEEE 802.11 standards is specifically devoted to vehicular environments~\cite{WAVE2}.
The two standards serve different purposes. Indeed, while some vehicular services are better served through a cellular infrastructure, others are more naturally mapped onto opportunistic communication schemes, especially those that are based on vehicle location and require extremely low latency. 

Several opportunistic communication paradigms for the local dissemination of contextualized information have been proposed. Relevant examples are Floating Content~\cite{floating,ours2017mobiworld,itc29,MONET2017,manzo2019analytical}, or Hovering Information~\cite{castro2009hovering,Nikolovski_2017}. They all aim at the local dissemination of information to end users over a defined geographic area (called Anchor Zone or AZ) through direct terminal-to-terminal connectivity. In what follows, we will refer to this approach as \emph{FC}. FC aims to minimize the usage of resources (such as bandwidth and memory) within the AZ, while achieving a given performance  (e.g., minimum message availability). Typically, FC performance at a given time instant is defined in terms of \textit{success ratio}, i.e., the average fraction of nodes with content at a given location (henceforth denoted as \textit{Zone of Interest} or ZOI).
FC paradigms find applications in several scenarios such as urban environments~\cite{fc_urban_env} and university campus contexts~\cite{campus}, for location-aware services.

Performance studies of FC schemes can be categorized into two classes. On one side, many works focus on content persistence over time and propose heuristics for guaranteeing content persistence within a predefined region, which usually coincides with the AZ itself~\cite{floating}. Such heuristics are often tailored to a specific context (e.g., highways or pedestrians in city squares~\cite{fc_urban}), and 
assume that a high likelihood of content persistence is sufficient to successfully support applications such as notifications of car accidents or traffic congestion. But they are hard to generalize for other applications that need a minimum amount of delivered contents within a given area.

Another class of proposals adopts a model-based approach to FC dimensioning in terms of AZ. Solutions in~\cite{floating,ours2017mobiworld,itc29,MONET2017,fc_urban_env,campus,ali2013impact,fc_urban} are based on the \textit{mean field approximation} of the dynamics of the population of users with content~\cite{mean_field}, and on the assumptions of stationarity of the mobility patterns and uniformity of user distribution in space. In these formulations, the FC dimensioning problem boils down to finding the minimum radius of a circular AZ area, which guarantees a given minimum content lifetime, and/or a target success ratio. However, models based on mean field approximation require a large enough population of users to yield accurate results, and this imposes a coarse granularity in the way in which resources are partitioned and managed. This leads to suboptimal AZ configurations, often including users who do not contribute significantly to FC performance, and hence large resource inefficiencies. 

Such approaches do not lend themselves to generalizations to more realistic settings characterized by heterogeneous user distributions (and in particular to user clustering, which is very common during traffic congestion or car accidents) and to settings in which mobility patterns vary significantly over time.
For these reasons, how to efficiently allocate resources (i.e., memory, bandwidth, and infrastructure support for content download) in an FC scheme with a realistic setting is still an open issue to date.

%

In this work, we consider a setup where infrastructure support to FC (in the form of a collection of data on user mobility, and orchestration of content replication) is ubiquitous. We propose \emph{DeepFloat}, a data-based approach for dynamic management of FC in vehicular scenarios that minimizes the overall use of bandwidth and of user storage space, as well as the amount of content seeding (i.e., content downloads) performed by the infrastructure over the target content lifetime. Our approach adopts a Convolutional Neural Network (CNN) architecture, typically used in image processing, in order to effectively capture the complex relations between elements of a multidimensional set of system parameters, their impact on content diffusion and persistence, and the FC success ratio. Different from existing results, our approach applies to arbitrary mobility scenarios and spatio-temporal patterns of vehicle density distributions, by proactively modulating the content replication and storage strategies over time. We use a configurable cost function to flexibly account for heterogeneity in vehicle populations (in terms of resource availability) and different resource cost models, as well as for the variations of these parameters over time and space. In particular, and to the best of our knowledge, our approach is the first to optimally adapt the balance between opportunistic replications (i.e., vehicular-to-vehicular communication) and direct content delivery by the infrastructure to the specific context.

The numerical assessment on a set of realistic scenarios shows very high levels of accuracy of our system, while substantially decreasing resource costs, with respect to model-based strategies. In particular, in contexts of rapidly varying mobility patterns (in which traditional approaches fail in delivering the target performance and in ensuring content persistence), we show its ability to proactively adapt content replication and availability over time in order to minimize the amount of infrastructure support to content replication in a QoS aware manner.

The paper is structured as follows. Section~\ref{system} presents the system model, followed by the problem formulation in Section~\ref{prob}. Our deep learning algorithm is illustrated in Section~\ref{alg} and assessed numerically in Section~\ref{N-e}. Finally, Section~\ref{concl} concludes the paper. 
\section{System Model}
\label{system}

We consider a set of wireless nodes moving on the plane according to a given road grid. These nodes are modeling as vehicles and/or pedestrians in an urban area with their user equipment (UE). We assume that the area is served by a cellular access network, which collects data about users mobility (\textit{floating user/car data}~\cite{floating_car}) for optimizing its operations, consistently with the 5G (and beyond) paradigm. 
We assume that each UE  is equipped with one or more wireless technologies for direct exchange of information with other UEs via opportunistic communications, such as WiFi Direct, Bluetooth, or device-to-device communication. 

We assume that the surface of the road grid to be partitioned into a set of L \textit{road links} (henceforth simply denoted as \textit{links}). 
The number and the shape of each road link is based on the tradeoff between computational complexity and accuracy of our approach described in Section~\ref{alg}.

We assume that two nodes are in \textit{contact} when they are in range to exchange information. We assume that at each time instant, the system maintains, for each road link, a list of contacts, e.g., by having nodes periodically exchange beacons. At any time, each node knows its exact position in space (i.e., its GPS coordinates). 

\subsection{DeepFloat FC operation}

We assume that at some point in time a content object (e.g., a piece of text, a picture, or a short movie) is generated either by the infrastructure (e.g., an advertisement from cellular or WiFi users), or by mobile nodes themselves (e.g., a warning about a road accident).

We consider a finite time interval, the \textit{floating period}, corresponding to the period during which the content should be made available to users in a target portion of the road grid (the \textit{Zone of Interest} or ZOI). The typical duration of the floating period ranges from half an hour up to a whole day. We assume that the floating  period is partitioned into $T$ intervals, each of duration $d_t$, $t=1,..,T$. At the beginning of the floating period, we assume that the infrastructure possesses some aggregate information about node mobility. This assumption is realistic as pedestrian and vehicular mobility patterns in urban scenarios exhibit predictable (and often periodic) behaviors over time~\cite{mobility_pred}.

An FC scheme is identified, for every road link $l\in \{1,...,L\}$ and time interval $t\in\{1,...,T\}$, by the values of \textit{content infectivity } $a_{l,t}$, of \textit{ recovery rate } $b_{l,t}$, and of \textit{seeding ratio } $s_{l,t}$ (with $a_{l,t}, b_{l,t}, s_{l,t} \in[0,1]$). 

At the beginning of each time interval $t$, the system makes sure that in each link $l$ the percentage of nodes possessing the given content at that time (henceforth denoted as content \textit{availability} at that time) is exactly equal to $s_{l,t}$.
At the beginning of each interval $t$, if in each link $l$ the content availability is less than $s_{l,t}$, the system transfers the content object from the infrastructure to a number of users sufficient to achieve the target value of availability. If the initial availability is larger than the seeding ratio for that interval, the system triggers an appropriate number of users to drop the content.

In the FC scheme, at the beginning of the floating period, parameters $a_{l,t}, b_{l,t}$ are made available to each user (e.g., broadcast to all users in the area through the cellular infrastructure). In each time interval $t$, when two nodes come into contact, if the content object is present at only one of the two nodes, the node with content object transfers it to the other node with probability $a_{l,t}$, where $l$ is the link where the sender node resides. Such transfer is subject to all limitations of the communication capacity between the two nodes, which is influenced by propagation effects such as fading and interference, among others.
When a content object is successfully transferred, the receiver (residing on link $l'$) keeps it with probability $b_{l',t}$, and discards it otherwise. 
Moreover, for all links $l$, whenever a node enters link $l$ with content, it keeps it with probability $b_{l,t}$ and drops it otherwise.

In this paper, we consider the ideal case in which nodes do not replicate content when it is not needed (i.e., when both nodes in contact already possess the content). We also assume that content exchanges are always unicast (one-to-one). However, this FC scheme can be easily extended to include the effects of multicasting and broadcasting. 

The resulting array $\mathbf{A}=\{a_{l,t},b_{l,t},s_{l,t}\}$ of dimensions $L \times T$ completely describes an FC scheme. Each entry in $\mathbf{A}$ is a 3-tuple identifying the content seeding, replication and caching strategies over all links during the whole floating period. 

In each interval, if enough content replications take place after the initial seeding, the content persists over time in a portion of the road grid even when the nodes which received the initial seeding have moved out of the road grid, or discarded the content. We say in this case that the content \textit{floats}, i.e., it persists probabilistically in the area. This usually happens for a duration which is determined by the mobility patterns, and the choice of the FC parameters $\mathbf{A}$, among others.

As a consequence, the primary performance metric for FC is the mean content \textit{availability}, i.e., the mean fraction of users with content over a given time interval. 
Indeed, as mentioned, the goal of an FC scheme is to make the content available in a subset $\mathcal{L}'$ of links in the Zone of Interest (ZOI) of the content, with a given mean availability, which we henceforth name as \textit{success ratio} and denote with $\alpha_t$. More specifically, an FC scheme aims at ensuring that, in every interval $t$, the condition  
\vspace{-0.2in}
\be\label{eq:cond1}
\alpha_t=\frac{\sum_{l\in\mathcal{L'}}n^c_{l,t}}{\sum_{l\in\mathcal{L'}}n_{l,t}}\geq \alpha_0
\ee
is satisfied, where $n_{l,t}$ ($n^c_{l,t}$) is the mean number of users (resp. mean number of users with content) in the road link $l$ at interval $t$, and $\alpha_0$ is the target minimum value of success ratio. The value of $\alpha_0$, the choice of size, shape, and location of the ZOI depend entirely on application requirements.

 \section{Formulation of the optimization problem}
\label{prob}
A cost function models the impact of an FC scheme on the number of employed resources. For each link and time interval, such function is given by the sum of three components. The first component accounts for the utilization of storage resources. For each link and time interval, it is measured by the product of the mean total number of users with content $n^c_{l,t}$ and the content size in bits, $D$.

The second cost component accounts for the mean amount of UE communication resources used to exchange content and it is proportional to  
the mean total number of UEs transmitting the content at a given time instant.
Specifically, let $\mathcal{U}$ be the set of communication technologies available (e.g., Bluetooth, WIFi direct, D2D, LiFI), such component is given by
\[
D \sum_{l,t}\sum_{u\in \mathcal{U}} \theta_{l,t,u}\gamma_{l,t,u}
\]
where $\gamma_{l,t,u}$ is the mean total number of UEs transmitting the content at time interval $t$ and link $l$ with the $u$-th communication technology. The coefficient $\theta_{l,t,u}$ can be used for adjusting the cost of opportunistic content exchanges across technologies, road links and time intervals, in order to take into account the availability of resources in that technology (e.g., due to utilization, to interference levels), and of the way  it varies over time. In general, $\gamma_{l,t,u}$ depends on content size, available bandwidth, and distance between transmitter and receiver. 

The third cost component takes into account the number of content transfers, implemented directly through the infrastructure (e.g., through the cellular network), for seeding the content in those links in which the availability at the beginning of a given time interval is inferior to the minimum content availability in the given road link. Hence, it is given by
\[
D \sum_{l,t} \left[s_{l,t}-v_{l,t}\right]^+
\]
(where $\forall x,\;[x]^+$ stands for $\max(x,0)$), and it is based on the assumption that content suppression ``comes for free". $v_{l,t}$ is the availability at the beginning of interval t, \textit{before} content seeding. It is given by
\vspace{-0.2in}
\be\label{eq:S_multicont}
    v_{l,t} = \left\{\begin{array}{lr}
    0 & t=1\\
    \frac{n^c_{l,t-1}}{n_{l,t-1}} & t>1 
    \end{array}\right.
\ee
An optimal FC scheme $\mathbf{A}$ is hence a solution to the following problem:

\begin{problem}[FC resource optimization]\label{prob:1}
\begin{equation}\label{eq:objectivefunction}
\min_{\mathbf{A}}\;\;\sum_{l, t} \frac{d_t D\left(n^c_{l,t}+\beta\sum_{u\in \mathcal{U}} \theta_{l,t,u}\gamma_{l,t,u}\right)}{\sum_{t=1}^Td_t} +\delta D \left[s_{l,t}-v_{l,t}\right]^+
\end{equation}
	\vspace{-0.15in}
	\begin{align}
	&\forall t,\gap\gap \alpha_t\geq \alpha_0\label{eq:availability_zoi}\\
	&\forall t, l\gap\gap 0\leq a_{l,t}\leq 1,\;\;\gap\gap 0\leq b_{l,t}\leq 1, \gap\gap 0\leq s_{l,t}\leq 1
	\end{align}
\end{problem}

The overall cost function is given by storage, opportunistic content exchange, and content seeding components. Those components are weighted by the ratio between the interval duration and the duration of the whole floating period, $\sum_{t=1}^Td_t$. Coefficients $\beta,\delta\geq 0$ modulate the relative weight of the cost components (memory, bandwidth, and seeding) on the overall cost of an FC scheme. $\theta_{l,t,u}$, which modulates technology and interference costs, together with $\beta,\delta$ are the instruments by which network operators can orchestrate the 
Floating Content service.  
Among all the possible FC schemes $\mathbf{A}$, a special role is played by the one in which $\forall \;l,t,\;a_{l,t}=b_{l,t}=s_{l,t}=1$. With such choice, the content is replicated at every opportunity over the whole road grid, and it is never dropped. This scheme, which we call \textit{all-on}, allocates all user resources in the system to the FC communication scheme and guarantees that at the beginning of each interval, each user possesses the content. This implies that for each time interval, the all-on scheme achieves the highest possible value of success ratio in the considered scenario. Therefore, if the all-on scheme is not a feasible solution of Problem 1, the problem is infeasible. When this is the case, Problem 1 has to be reformulated by providing more resources to FC, e.g., by considering a larger road grid, by including a larger population of users, by changing $\alpha_0$ or communication technologies.

In general, Problem $1$ cannot be solved efficiently. Indeed, in general $\forall\;l,t,\;n^c_{l,t}$ and $\gamma_{l,t,u}$ depend on $\mathbf{A}$ as well as on a large set of  parameters, in ways which are hard to accurately capture analytically without strong and unrealistic assumptions, such as stationarity of mobility patterns and uniformity of user distributions. In the next section, we present our approach to this intractable problem, based on Deep Learning.
\section{A deep learning algorithm for efficient FC dimensioning}
\label{alg}
%
In this section, we illustrate our DeepFloat approach to finding an efficient solution of Problem~\ref{prob:1}.
It is based on Convolutional Neural Networks (CNN), a specialized type of neural networks that has proven particularly effective in modeling data with grid-like topology \cite{goodfellow2016deep}. The key idea of our approach is to exploit this CNN property in order to accurately and efficiently model the correlation between the spatial features of the road grid, and the way in which the content spreads among links and persists over time. Our approach is divided into three phases:

\begin{itemize}
    \item \textbf{Offline Set Up}. The cellular infrastructure collects and records UE mobility traces over time 
    across the whole road grid, on a regular basis. The training set is generated, by associating the communication features to the measured mobility features. Finally, the CNN is trained. 
    \item \textbf{Bootstrap}. It starts when an application (which may or may not reside within a UE) requests to the system, via the cellular infrastructure, the activation of the FC scheme, with a given ZOI, target success ratio, and floating period. The system, given the forecasted resource availability and mobility patterns, uses the trained CNN to compute in real time a strategy $\mathbf{A}^*$ that achieves the target success ratio $\alpha_0$.
    \item \textbf{Deployment}. The system provides to all UEs present in the considered road grid at the beginning of the floating period (as well as to all those UEs which enter the road grid during the floating period) the coefficients $a^*_{l,t}$ and $b^*_{l,t}$, which identify the replication and caching strategy derived by the CNN. At the beginning of each time interval, in each link, the content is seeded according to $s^*_{l,t}$. If necessary (i.e., if, during the floating period, mobility patterns in the road grid deviate significantly from the forecasted ones), new forecasts are elaborated, and a new strategy $\mathbf{A}^{**}$ is adopted for the rest of the floating period.
\end{itemize}


In what follows, we describe in detail the three phases, and the algorithms involved. 
\subsection{Offline Set Up}


\subsubsection{Data collection}
As we have mentioned, the system collects, on a regular basis, data about UEs mobility in the grid. The system divides content lifetime into intervals to provide a more efficient strategy that follows UEs mobility. For each time interval, the system records the trajectories of each UE in the grid. 

Starting from these trajectories, for each interval $i$, and every link $l$, the system computes a set of aggregate metrics relative to node mobility and to wireless communications. In what follows, these metrics are the average node speed, the average number of nodes, and average contact duration. In addition, it computes the mean number of nodes that, in a given time instant, are in contact (i.e., able to exchange beacons) with a node. Notice that nodes of different road links can be in contact.  
These parameters have been chosen as they are typically used as input in the main existing analytical models of FC~\cite{ours2017mobiworld,itc29}. Of course, however, different choices are possible. In general, the choice of the set of parameters, as well as the aggregation level in space (i.e., size and shape of links) and time (i.e., duration of an interval $d$) affects the degree of accuracy with which our model accounts for the spatio-temporal features of the system, and specifically, for the mobility patterns. 
The mobility parameters values of the whole map, for each interval $i$, are stored in a \textit{mobility feature array} $\mathbf{m}_i$.

\subsubsection{Label generation}
In the next step, called \textit{label generation}, a randomization procedure is applied. Specifically,

\begin{itemize}
\item 
for each $\mathbf{m}_i$, a set of $K$ random FC schemes $\mathbf{A}_k$, $k=1,...,N$ are generated, each with its seeding strategy. 
\item For each $i$ and each pair $(\mathbf{m}_i,\mathbf{A}_k)$, a simulation is performed based on the random strategy $\mathbf{A}_k$ and the user trajectories during interval $i$. For each communication technology, these simulations assume a given model for the channel capacity between nodes over time. The parameters measured, for each link, are the mean number of nodes with content at a given time instant $n^c_{l,i}$, and the mean number of nodes that are transmitting at a given time instant for each communication technology, $\gamma_{l,i,u}$. These parameters constitute the \textit{communication feature vector}, $\mathbf{c}_i$. Given a choice of ZOI, these parameters are the basis for the computation of the success ratio, as well as for the estimation of the resource utilization and of the associated costs. However, note that the knowledge of the ZOI or of the target success ratio is not required to produce the training set. Hence this phase can be performed entirely offline.
\end{itemize}

For each k and i, with $\mathbf{P}_i=(\mathbf{m}_i,\mathbf{c}_i)$ we denote the \textit{Link Features Vector} associated with $\mathbf{A}_k$ (Table~\ref{Tab1:features}). Finally, each vector $\mathbf{P}_i$ is normalized, in order to avoid numerical issues in the subsequent phases of the process.
The output of the offline phase is thus a set of pairs $(\mathbf{P}_i,\mathbf{A}_k)$, which we denote with $\Pi$. 
The set $\Pi$ is therefore enriched over time with new elements derived by simulations. In addition to those elements, the system includes in $\Pi$ also those elements derived by directly measuring FC performance (specifically, those in the communication feature vector) during the operation of the FC schemes. Note that, while the presence of such measured elements helps in achieving a high level of accuracy in real scenarios, the contribution to $\Pi$ of those elements derived through simulations plays a key role in enabling the system to appropriately configure an FC scheme in those scenarios which occur with a relatively low probability (e.g., road accidents, sport events, or disasters). 

\begin{table}[t]
\caption{The set of link features for time interval $i$ and link $l$.}
\centering
\small
\begin{tabular}{  c l }
 \hline
 \textbf{Name} &   \textbf{Description} \\
 \hline
 
 $n^c_{l,i}$ & average number of nodes with content\\
 
 $n_{l,i}$ &  average number of nodes \\
 
 $\lambda_{l,i}$ & average number of nodes in contact\\
 
 $\tau_{l,i}$ & average contact duration $[s]$ \\
 
 $\nu_{l,i}$ &  average speed $[m/s]$\\
 
  $\gamma_{l,i,u}$ & average number of nodes transmitting\\
  &  in the $u-th$ technology\\
 \hline
\end{tabular}
\label{Tab1:features}
\end{table}
\vspace*{-0.1cm}

\subsection{Bootstrap and Deployment}
The bootstrap phase starts when a request for FC service is made. A request is characterized by an indication of the region of the road grid which constitutes the ZOI, by the target value of success ratio $\alpha_0$, and by the floating period.

In what follows we assume the floating period to be composed by $T$ intervals, and let $\mathbf{m}_t$ be the mobility feature array for the $t-th$ interval, $t=1,...,T$. For all intervals $t$, we assume the arrays $\mathbf{m}_t$ to be perfectly known at the time of the service request, possibly by forecasts based on the mobility traces which the system collects on an ongoing basis (we will later describe how our approach accounts for forecast errors). \footnote{The mechanism by which these arrays are forecasted is out of the scope of the present paper}.
Given $\mathbf{M}=\mathbf{m}_1, ...,\mathbf{m}_T$ the forecasted mobility features for the floating period $T$, the definition of the ZOI (i.e., the set $\mathcal{L'}$), and the target success ratio, the CNN computes the replication, caching, and seeding (i.e., the vehicles downloading the content from the infrastructure) strategies  $\mathbf{A}^*=\mathbf{A}^*_1,...,\mathbf{A}^*_T$ for the whole floating period, which achieve the target success ratio while trying to minimize the resource cost as defined in \eref{eq:objectivefunction}. For these strategies, the content infectivity (i.e., replication probability), as well as the recovery rate (i.e., the caching probability) for each link and time interval within the floating period are communicated to all nodes in the scenario. In those cases in which, during the operation of an FC scheme, better forecasts of mobility become available, the system computes a new strategy $\mathbf{A}^{**}$ for the remaining portion of the floating period, based on such forecasts, and it injects the new values of content infectivity and recovery rates to all the nodes. 
  
\subsection{CNN Model Architecture}
\label{sec:CNN}
In this section we describe the architecture of the CNN adopted in our approach.

In the FC dimensioning problem ---definition of the replication, caching and seeding strategies--- the information about proximity or relative position between links is not part of the link features. 
In general, the correlations between features of different road links are not due to proximity only. Rather, they are also the result of wireless propagation effects and, most importantly, of spatio-temporal patterns in node mobility (i.e., typical patterns of vehicular traffic). 

The Convolutional Neural Network (CNN) architecture enables capturing both intra-link and inter-link relations between features. The features extracted by our CNN are associated with the correlation between road links. 
As a result, the characteristics extracted from each layer of the CNN do not consist only in local correlations, but also in long-spatio-temporal relations between links, which have a strong impact on FC performance. 

\begin{figure}[t!]
 \begin{center}
    \includegraphics[width=\linewidth]{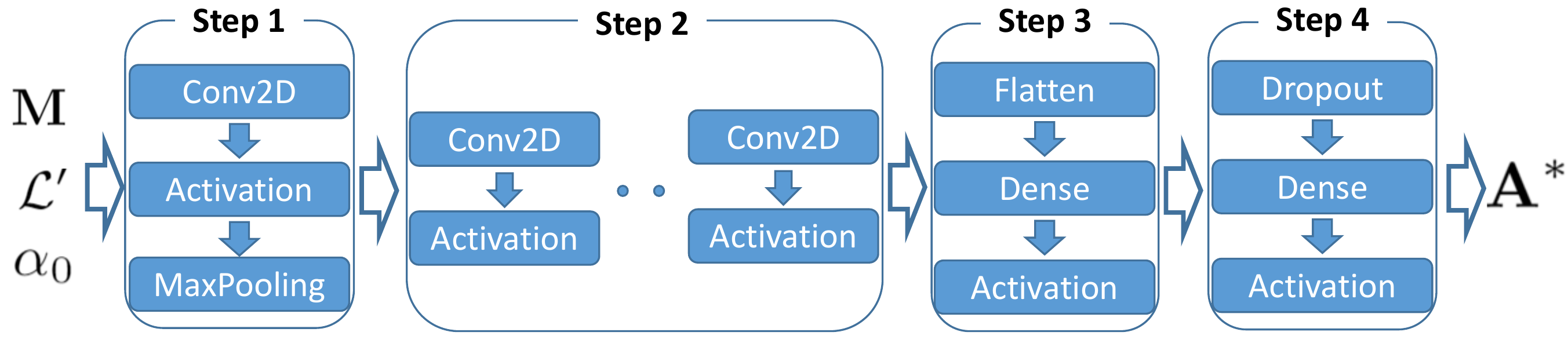}
\caption{\small Outline of the architecture of our Convolutional Neural Network. It takes as input the mobility features $\mathbf{M}$, as well the set of links composing the ZOI, $\mathcal{L'}$, and the target success ratio $\alpha_0$. Its output are the replication, caching, and seeding strategies $\mathbf{A}^*$.}
\label{fig:cnn_arc}
\end{center}
\normalsize
\vspace{-0.2in}
\end{figure}

The detailed structure of the CNN architecture consists of four steps (\fref{fig:cnn_arc}). 
The goal of step 1 is to learn features corresponding to basic patterns in the input (e.g., the strongest local correlations, such as the spatio-temporal correlations between adjacent road links and between the features of the same road link.)~\cite{goodfellow2016deep}. 
Step 1 consists in a convolution layer, which filters the input data (layer \textit{Conv2D} in \fref{fig:cnn_arc}). The size of the convolution kernel determines the extent of the locality of the correlations captured. The output is fed into an \textit{activation layer}, followed by a discretization process (\textit{max pooling} layer). The purpose of these two layers is to filter out information which will not be relevant for subsequent parameter optimization.

In step 2, to improve model accuracy and to avoid overfitting, the same operations as step 1 is repeated. Note that, in step 2 the max-polling layer is unnecessary given the data structure.  
The number of these layers directly defines the depth of the CNN and the level of complexity of the correlations able to capture. Hence, their number depends on the characteristics of the modeled system, but also on a tradeoff between computational complexity and accuracy of the output. The choice of kernel size in the convolutional layers (Conv2D) should also be done according to such tradeoff. Indeed, a small kernel forces the use of several convolution steps in order to capture long-range correlations (i.e., correlation between road links). A large kernel instead decreases the amount of Conv2D layers required in step 2, but it also prevents the CNN from accurately modeling complex correlations.
The purpose of step 3 is to reshape the output of the previous step, in order to facilitate the subsequent computations.
Its output is a set of candidate strategies for orchestrating DeepFloat system. Finally, step 4 performs an optimization over these candidate strategies, by selecting out iteratively those who do not satisfy the performance constraint on success ratio, and by giving as output the one which minimizes the cost function in \eref{eq:objectivefunction}. 

\subsection{Complexity}
A crucial aspect of the performance in our approach is the computational load required by the three phases which compose it.
Given the potentially large amount of data to be collected and pre-processed, the offline set up is the most computationally intensive phase, being directly related to the size of the set $\Pi$, as well as to the number of links and features considered. However, given that all the elaborations in the offline setup phase are executed offline, they do not impact the delay with which a request for Floating Content is served.

The computations required by the bootstrap phase are performed in real time at the moment a request is made. Therefore, those effects heavily impact the overall FC performance, particularly for those applications for which a quick and effective deployment of the FC scheme is often necessary such as car accident warning or medical emergency notification. In order to enable a fast deployment in these scenarios, several ways exist to decrease the computational load at the bootstrap time within our approach. A possibility, as discussed, is to adopt a low number of blocks at step 2 in the CNN.

\section{Numerical Evaluation}
\label{N-e}
In this section, we numerically assess the performance of DeepFloat, in terms of model accuracy and resource efficiency, and we characterize the spatio-temporal strategies emerging from it. 

\subsection{System Setup}
We assumed nodes using a single wireless communication technology, with a path loss attenuation of $3$ (typical of urban scenarios), a SINR value equal to $5$ dB (which models setups with high interference, common in crowded city centers, e.g., at the $2.4$ GHz frequency), content size $D = 8$ MB (to emulate the exchange of multimedia content such as pictures and short videos), bandwidth $B = 1$ MHz (as in Bluetooth) and the channel capacity resulting from Shannon's formula.  
Unless otherwise specified, the FC parameters and those of the CNN have been chosen conservatively in order to fit a worst case scenario, in which applications require a fast bootstrap phase, a relatively short content floating period ($1h$), and a high level of target success ratio ($0.9$). Both coefficients $~\beta$ and $\delta$ in the cost function \eref{eq:objectivefunction} are set to 1 (to have replication, caching, and seeding equally contributing). As we show in this section, a CNN architecture with a single block at step 2 and a Conv2D kernel of size $3$-by-$3$, together with the seeding approximation described in \sref{sec:CNN} have proven sufficient to achieve a good level of accuracy while minimizing computational load. In order to avoid over-fitting, a standard 10-fold cross-validation is performed.   

\subsection{Manhattan Scenario}

In the first set of experiments, we evaluated the performance of our proposed approach in a scenario with synthetic mobility, in order to perform, in a controlled scenario, a first-order characterization of the strategies emerging from the DeepFloat approach, and of their effectiveness and efficiency.

\begin{figure}[t]
 \begin{center}
    \includegraphics[width=0.5\linewidth]{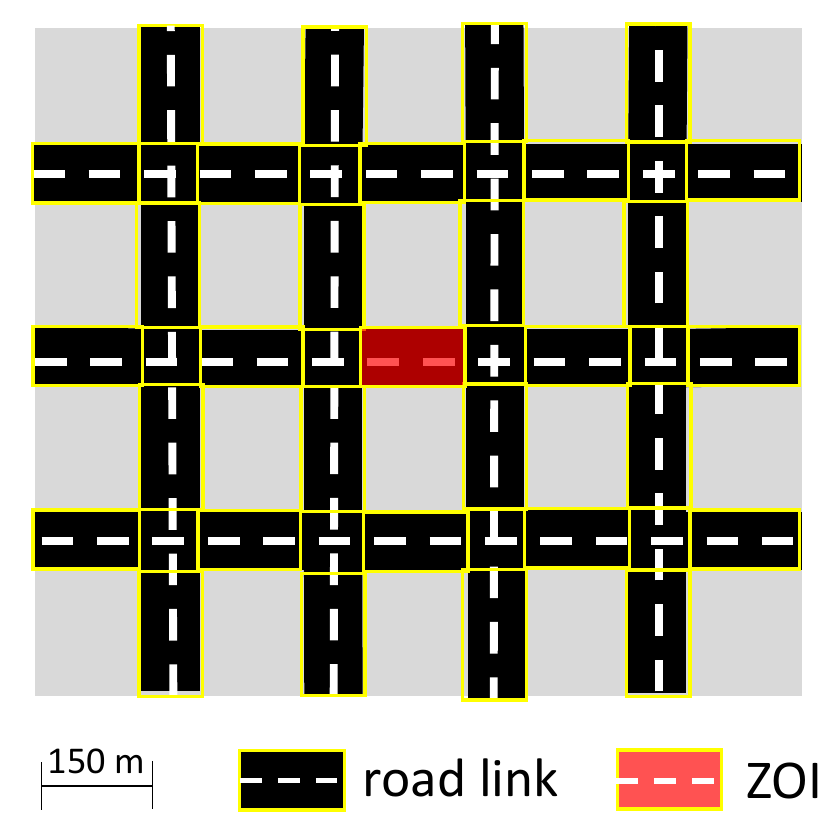}
\caption{\small Manhattan road grid composed of $5$-by-$4$ square blocks of side $150$ m, for a total of $31$ road links and $12$ road intersections.}
\label{fig:manha_grid}
\end{center}
\normalsize
\vspace{-0.2in}
\end{figure}

We considered the road grid in \fref{fig:manha_grid}. 
Nodes enter and exit the grid from road links at the border, with an arrival rate equal to $1.5$ nodes/s per road link, equal for all road links. At every crossroad, a new direction is chosen at random between straight, right and left. 
The floating period is composed by a single interval of 1h duration.
The training set was built as follows. The road link features were measured over $1h$, using a sampling interval of $1$s. In order to collect communication features, we considered $10^3$ different random strategies. 
The resulting simple randomization, in each scenario, contained about $4\cdot10^6$ pairs $(\mathbf{P},\mathbf{A})$. 
Only for the model accuracy study, we assume that mobility is predicted perfectly.

%
%
\subsubsection{Model Accuracy}
\begin{figure}[t]
 \begin{center}
    \includegraphics[width=0.9\linewidth]{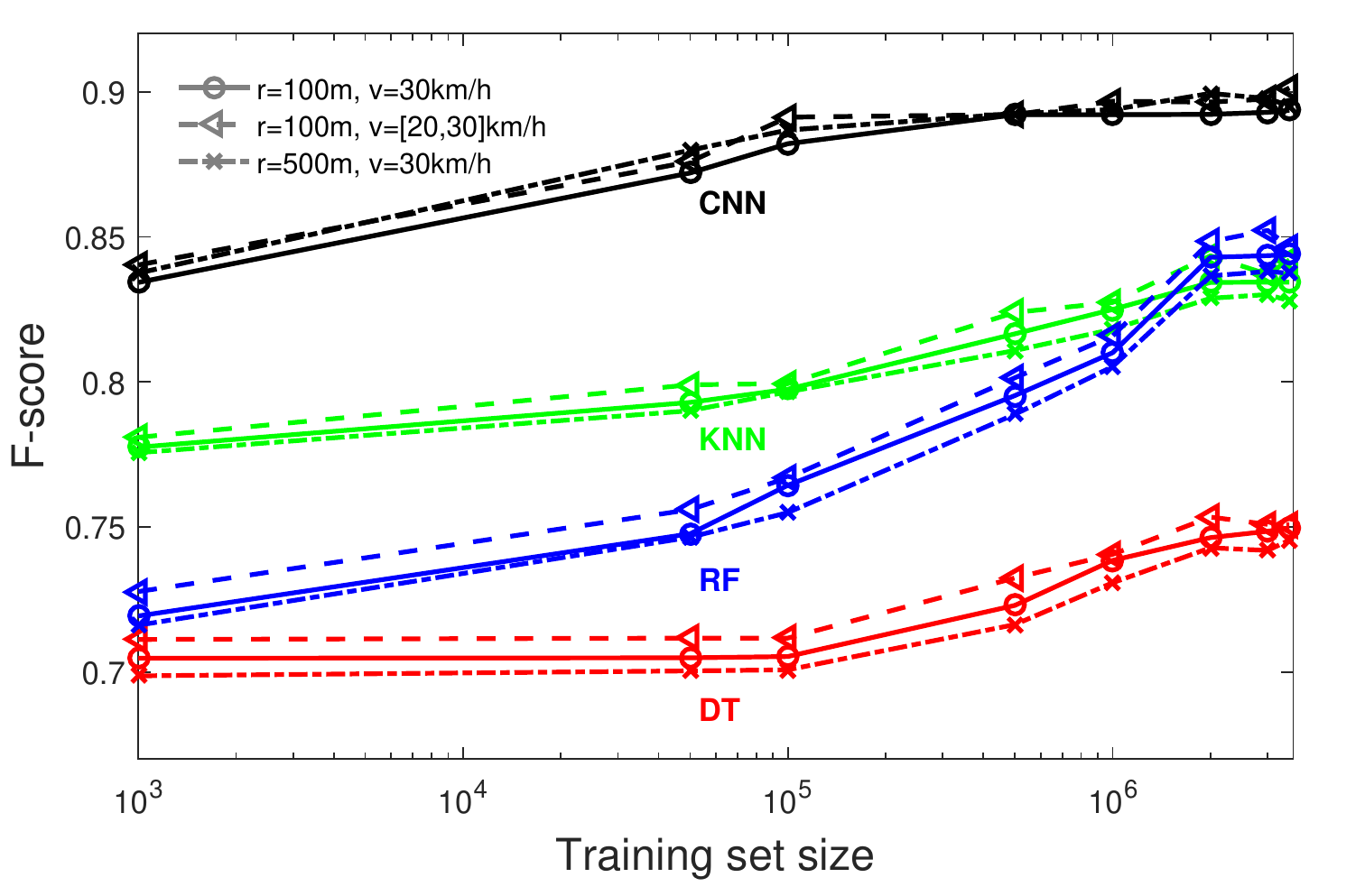}
\caption{\small F-score versus size of the training sample, for the DeepFloat approach (DF), as well as for K-Nearest Neighbor (KNN), Decision Tree (DT), and Random Forest (RF), for the three scenarios considered. All curves are with a $98\%$ confidence interval of $7\%$}
\label{fig:fscore}
\end{center}
\normalsize
\vspace{-0.2in}
\end{figure}

In order to evaluate the accuracy of DeepFloat, \fref{fig:fscore} shows the values of F-score~\cite{f-score} versus the training set size, for three different combinations of transmission radius and speed: $r = 100$ m and speed $\nu = 30$ km/h; $r = 100$ m and $\nu$ uniformly distributed in $(20, 30)$ km/h ---emulating vehicles queuing; $r = 500$ m and $\nu = 30$ km/h.
In each scenario, we also computed the performance of other algorithms for multi-labeled classification problems, i.e., K-Nearest Neighbor (KNN), Decision Tree (DT), and Random Forest (RF). Note that for each algorithm, all scenarios are tested using over $1.5\cdot 10^5$ registers of the test set, obtaining a high sample population for the confidence interval.

Results in \fref{fig:fscore} show that DeepFloat substantially outperforms the other algorithms in terms of accuracy. In particular, our approach achieves high accuracy even with small training sets. This confirms that for the problem of efficient dimensioning of an FC scheme, the specific ability of CNNs to capture the inter-link correlations is key for satisfactory accuracy performance. This is also suggested by the fact that, by increasing the transmission radius (and hence the amount of inter-link content exchanges), the relative accuracy of our approach improves. 

A key aspect of any learning approach to FC dimensioning is that even approaches with high accuracy may produce configurations that are unfeasible, i.e., which do not achieve the target performance. Table~\ref{tab:output} shows, for all considered scenarios and algorithms, the probability that the output of the learning processes does not satisfy constraint~\eref{eq:availability_zoi}. For our approach, the rejection probability is around $3\%$, an order of magnitude lower than for the other considered approaches. Indeed, CNNs tend to decrease the~\textit{false positive} predictions rather than the \textit{false negative} ones. This is due to the fact that, for any input, in the search for the strategy, which minimizes resource costs, our approach always considers the all-on configuration as a fallback option. 
\begin{table}[t!]
\footnotesize
\begin{center}
\caption{Rejection probability (i.e., probability of not achieving the target success ratio) of the four approaches considered. Values are with a $98\%$ confidence interval within $6\%$ of interval size. Training set size: $1.5\cdot10^5$.}
  \label{tab:output}
\begin{tabular}{  c c  c  c }
  \hline
    \textbf{Method}&$r=100$ m & $r=100$ m &$r=500$m\\
   &$\nu=30$ km/h&$\nu=[20,30]$ km/h&$\nu=30$ km/h\\
  \hline

  DF &\textbf{0.029} & \textbf{0.031} & \textbf{0.032}\\
 
  KNN &0.274 & 0.287& 0.284\\
 
  RF &0.324 & 0.320& 0.333\\
 
  DT &0.347 & 0.345& 0.349\\
  \hline

\end{tabular}
\end{center}
\normalsize
\vspace{-15pt}
\end{table}

\subsubsection{Resource Optimization}
\begin{figure}[t!]
 \begin{center}
   \includegraphics[width=.75\linewidth]{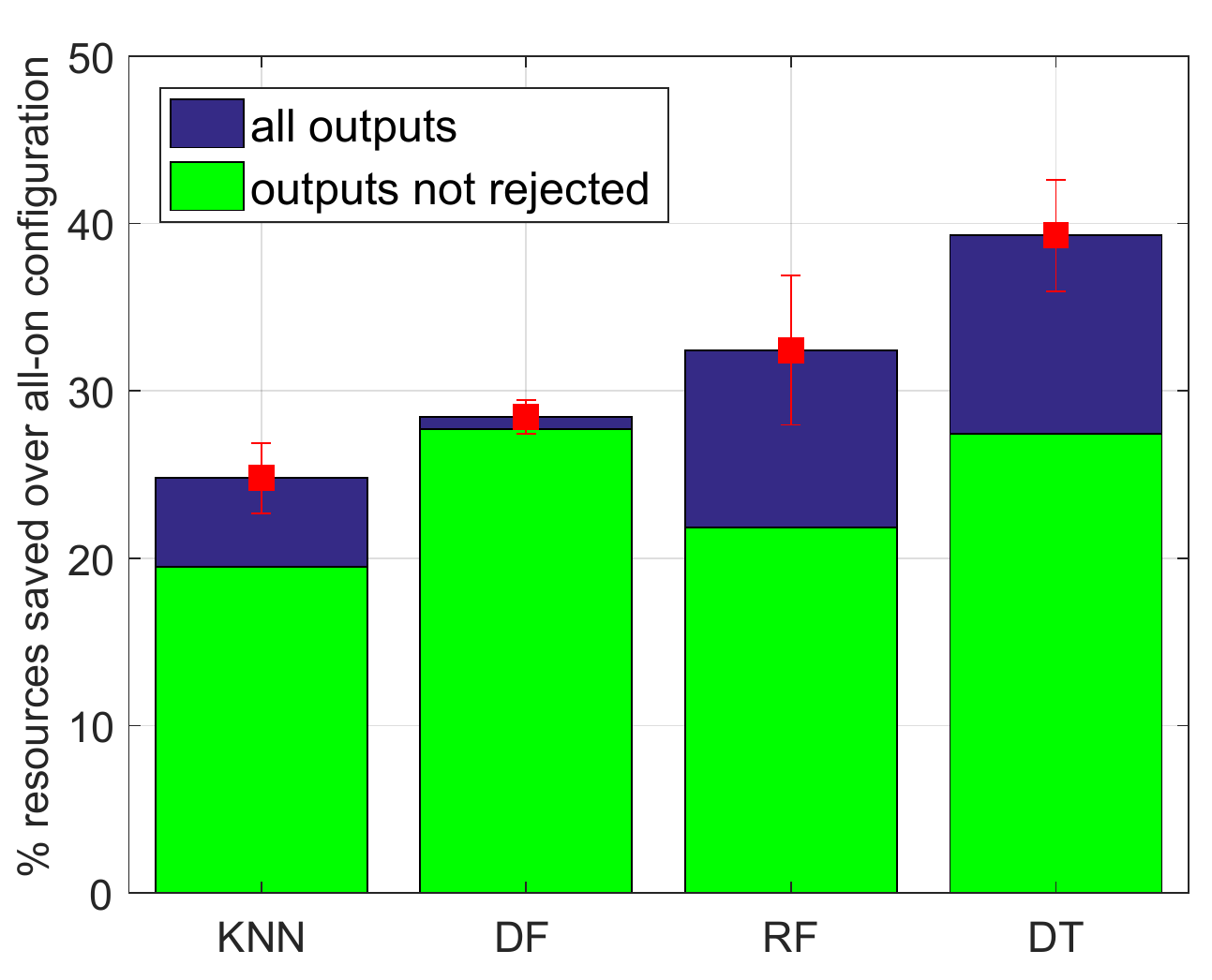}
\caption{\small Percentage of resources saved with respect to the all-on strategy, for the four approaches considered. $r=100m$ and $\nu=30km/h$. Values are with the confidence interval of $98\%$ with interval size of $3\%$.}
\label{fig:resources}
\end{center}
\vspace{-10pt}
\end{figure}

A crucial performance parameter of our CNN approach is the percentage of resources saved with respect to the all-on configuration. Such a  configuration is considered as a reference, as by hypothesis it is the only one that is always feasible. 
\fref{fig:resources} shows the percentage of resources saved with respect to the trivial case of the all-on configuration. 
They have been computed assuming that those configurations that do not achieve the target success ratio have the same cost of the all-on configuration. 
In Figure~\ref{fig:resources}, we present both the ideal case, where no predicted configuration has been rejected, and the case, which takes into account the rejected outputs. 
In the ideal case, by using the Decision Tree technique we save up to $39\%$ of the resources with respect to the all-on configuration. However, not all configurations resulting from DT satisfy the success ratio constraint. 
These results suggest that our CNN architecture is able to shape the content replication and storage strategies much more efficiently than classical, non-deep ML techniques. Indeed, our CNN is able to take advantage of the very high number of parameters used for training, achieving high accuracy without requiring a large training set or any geographical information (such as a road map).
%

\subsection{Luxembourg City Scenario}
\begin{figure}[t!]
 \begin{center}
    \includegraphics[width=0.75\linewidth]{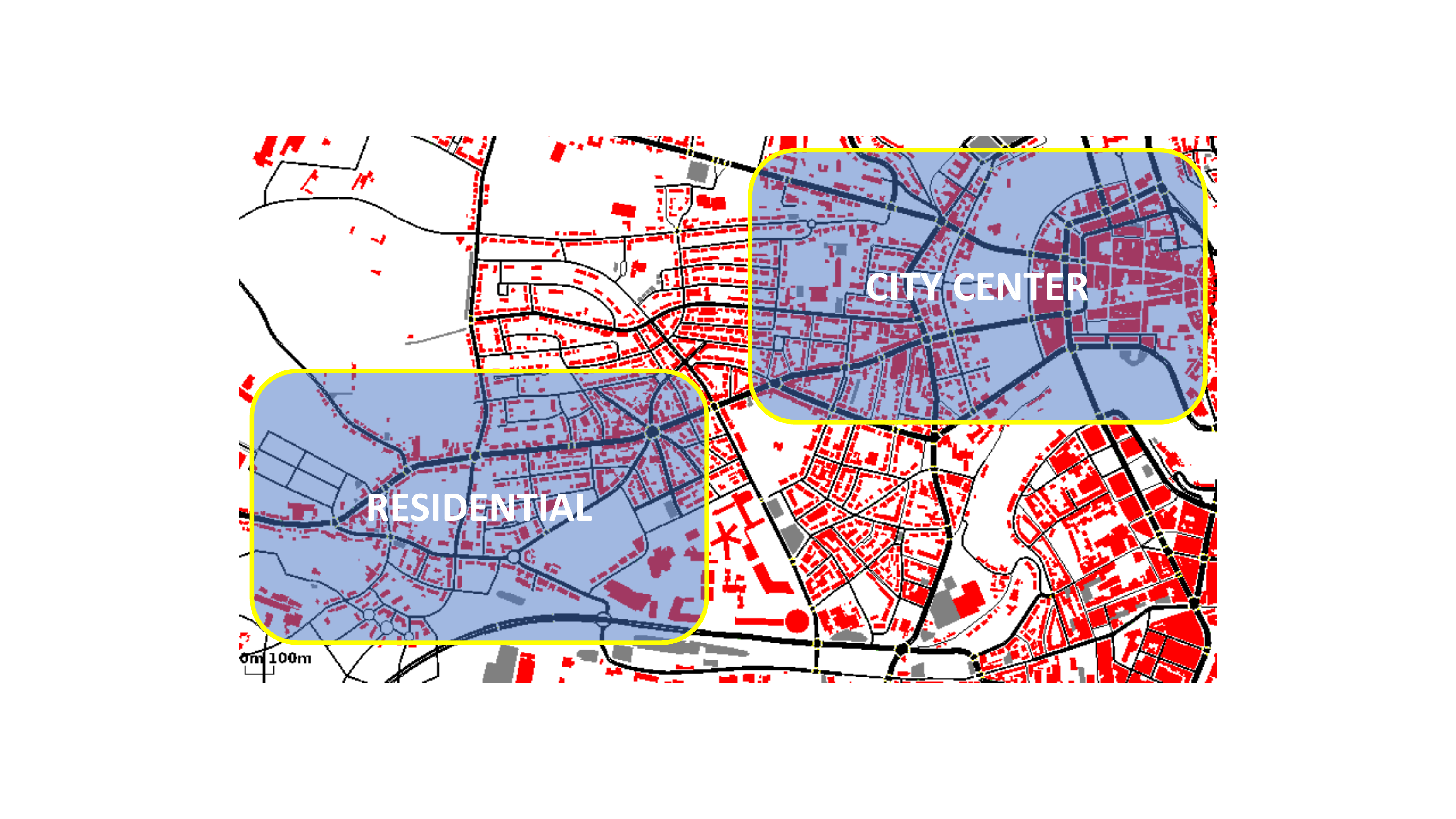}
\caption{\small Luxembourg City map, with the borders of the City center and Residential areas.}
\label{fig:lux_map}
\end{center}
\normalsize
\vspace{-10pt}
\end{figure}

In order to perform a more realistic assessment of the performance of our approach, we considered a second scenario, shown in \fref{fig:lux_map}. It consists of two districts (denoted as Residential and City center). In both districts, the street grid and the measurement-based mobility traces for the rush traffic hour (i.e., between 7 AM and 8 AM) were derived from \cite{lux}. 
These areas were selected because of their difference in vehicle density distribution. In the city center, during the rush hour, the vehicle density distribution is high and relatively uniform, while in the residential district it is lower and more clustered. 
In both scenarios, we considered as ZOI the closest link to the center of the district.  
In both districts, we collected data from $169$ road links, using a sampling rate of $1s$ to accurately capture mobility dynamics. 
The transmission radius is taken to be $100$m. Note that, Bluetooth Low Energy transmission range is greater than $100m$, whereas IEEE 802.11p (WAVE) reaches $1000m$~\cite{WAVE2}.

We assumed a floating period composed by a single interval of $1$ h, (adequate for a traffic warning e.g., about a temporarily restricted area). By choosing a single interval, we focus on the ability of the considered approaches to capture spatial correlations in the input data.
%

In Table~\ref{tab:real_accurary}, we list the test accuracy values obtained by using $1.5\cdot10^5$ registers for testing, and $10^6$ for training. While non-deep learning techniques have a worse accuracy in the Luxembourg scenario than in the Manhattan one, the contrary holds for DeepFloat, which is able to account for spatial inhomogeneities in the mobility features.

\begin{table}[t]
\footnotesize
\begin{center}
\caption{F-score for the Manhattan and Luxembourg scenarios, with a training set size of $10^6$, and test set size of $1.5\cdot10^3$.}
\label{tab:real_accurary}
\begin{tabular}{ c c c c c c}
 
   \hline
   \multicolumn{2}{c}{\textbf{Scenario}}&\textbf{DF}& \textbf{KNN}& \textbf{RF}& \textbf{DT}\\
   \hline

	\multirow{6}{*}{\rotatebox[origin=c]{90}{Manhattan}}&$r=100m$&\multirow{ 2}{*}{0.892}&\multirow{ 2}{*}{0.824}&\multirow{ 2}{*}{0.810}&\multirow{ 2}{*}{0.738}\\
   &$\nu=30 km/h$& & & &\\
   \cline{2-6}
   
   	&$r=100m$&\multirow{ 2}{*}{0.893}&\multirow{ 2}{*}{0.834}&\multirow{ 2}{*}{0.816}&\multirow{ 2}{*}{0.740}\\
   &$\nu=[20,30] km/h$& & & &\\
   \cline{2-6}
   
   	&$r=500m$&\multirow{ 2}{*}{0.894}&\multirow{ 2}{*}{0.819}&\multirow{ 2}{*}{0.802}&\multirow{ 2}{*}{0.736}\\
   &$\nu=30 km/h$& & & &\\
   \hline
   
   	\multirow{4}{*}{\rotatebox[origin=c]{90}{Lux.}}&\multirow{ 2}{*}{city center}&\multirow{ 2}{*}{\textbf{0.897}}&\multirow{ 2}{*}{0.802}&\multirow{ 2}{*}{0.800}&\multirow{ 2}{*}{0.726}\\
   & & & & &\\
   \cline{2-6}
   
   	&\multirow{ 2}{*}{residential}&\multirow{ 2}{*}{\textbf{0.896}}&\multirow{ 2}{*}{0.798}&\multirow{ 2}{*}{0.801}&\multirow{ 2}{*}{0.722}\\
   && & & &\\
   \hline
\end{tabular}
\end{center}
\normalsize
\vspace{-0.1in}
\end{table}

The main existing analytical result which addresses the issue of resource-optimal FC dimensioning in a road grid was presented in~\cite{ours2017mobiworld}. It assumes that content is replicated and stored within a circular area (called anchor zone - AZ) within which content infectivity and recovery rate are both set to 1. 

\begin{figure}[t]
 \begin{center}
    \includegraphics[width=.85\linewidth]{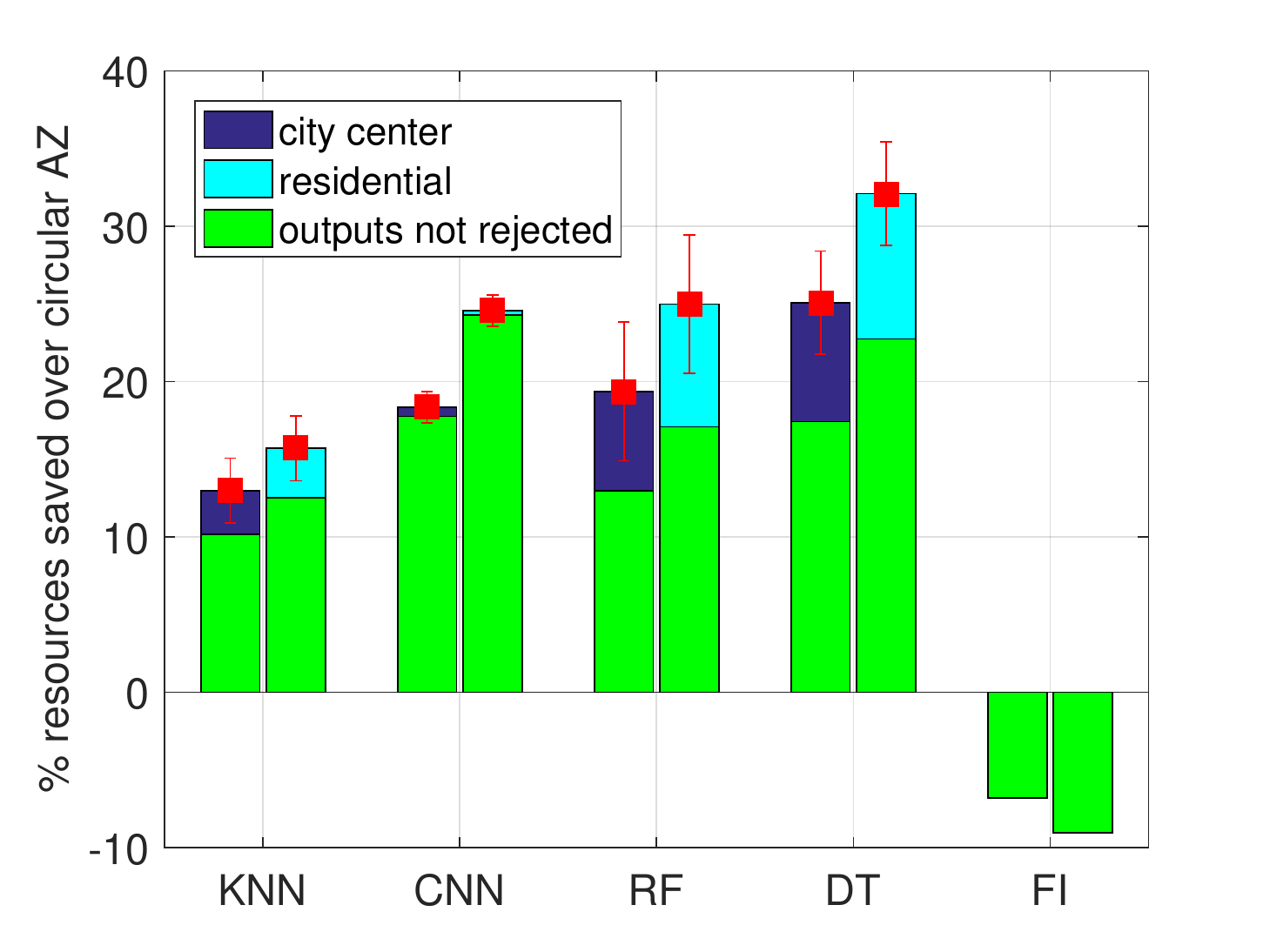}
\caption{ $\%$ of resource saved with respect to the Circular Anchor Zone approach~\cite{ours2017mobiworld}, for our  DeepFloat approach (DF), as well as for the K-Nearest Neighbor (KNN), Decision Tree (DT), Random Forest (RF) approaches, and Full-Infrastructure solution (FI) in the rush hour time interval. Confidence interval of $98\%$ with interval size of $3\%$.}
\label{fig:circular}
\end{center}
\normalsize
\end{figure}

For the two districts of Luxembourg City, we have evaluated the resource efficiency of our DeepFloat approach, as well as of the KNN, Decision Tree, Random Forest approaches, and Full-Infrastructure ---a base station providing the content object to the vehicles within the ZOI--- in terms of percentage of resources saved with respect to the approach from~\cite{ours2017mobiworld} (see Figure~\ref{fig:circular}). As in the Manhattan scenario, we have assumed that the cost of rejected outputs ---when the strategy does not satisfy the performance target--- is the same as the one of the all-on configuration. As we can see, all ML approaches outperform the analytical result. This is most likely due to the fact that the analytical approach is based on an assumption of uniformity of the main mobility features within the AZ. 
Indeed, the gains over the analytical result are larger in the less-uniformly distributed residential scenario than in the city center. 
%

Table~\ref{tab:output_rej_real} provides the output rejection probability for the Luxembourg scenarios. Similarly to the Manhattan scenario, the lowest values are provided by the DeepFloat approach in the city center ($1.8\%$). 

\begin{table}[t]
\footnotesize
\begin{center}
\caption{Rejection probability of the considered ML approaches. Values are with a $98\%$ confidence interval within $6\%$ of interval size. Luxembourg scenario.}
  \label{tab:output_rej_real}
\begin{tabular}{ c c c }
  \hline
  \textbf{Method}& \textbf{City Center} & \textbf{Residential}\\
  \hline
  DF &\textbf{0.018} & \textbf{0.021}\\
  
  KNN &0.224 & 0.229\\
  
  RF &0.307 & 0.314\\
  
  DT &0.332 & 0.341\\
  \hline
\end{tabular}
\end{center}
\normalsize
\vspace{-15pt}
\end{table}

\begin{figure}[t]
 \begin{center}
    \includegraphics[width=0.45\textwidth]{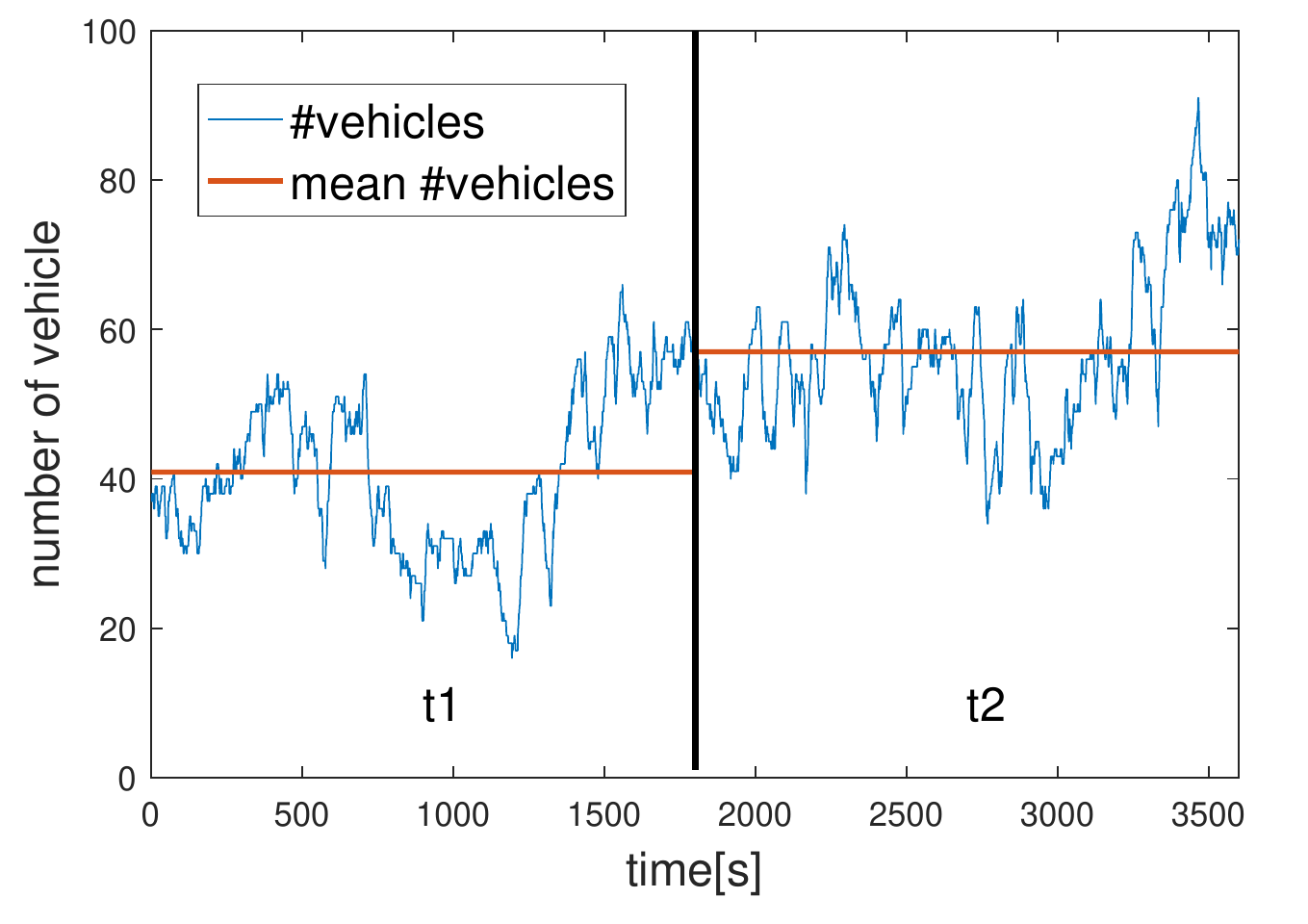}
   
\caption{Vehicle density over time in the Luxembourg city center scenario, for the first time interval $t1$ = (7 AM - 7:30 AM), and for the second, $t2$ = (7:30 AM - 8 AM).}
\label{fig:vehicle_density}
\end{center}
\normalsize
\end{figure}

\begin{figure}[ht]
 \begin{center}
    \includegraphics[width=0.45\textwidth]{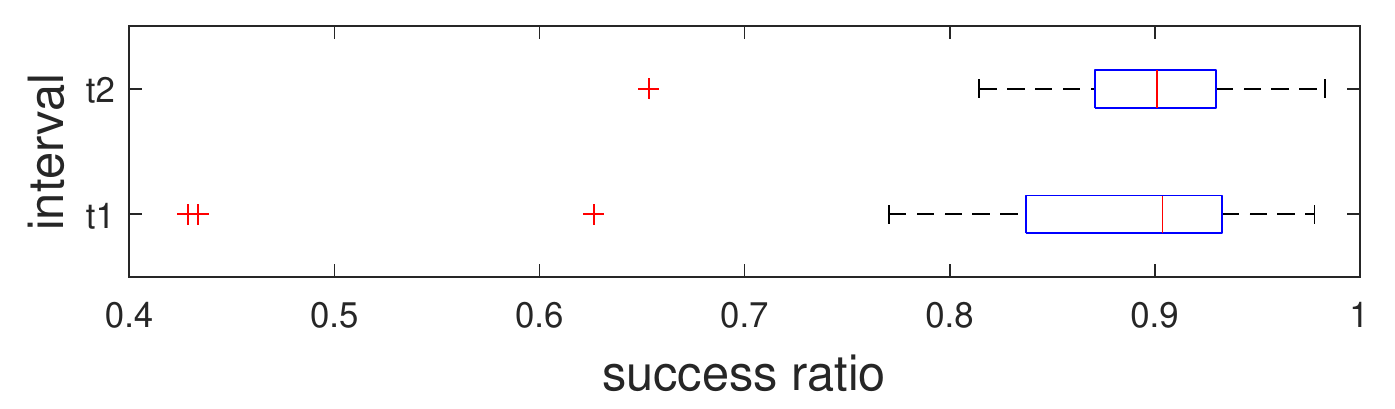}
\caption{Distribution of the measured success ratio derived with the strategies from the DeepFloat approach, in time interval t1 = (7 AM - 7:30 AM) and t2 = (7:30 AM - 8 AM), for the Luxembourg city center scenario.}
\label{fig:zoi_av_interv}
\end{center}
\normalsize
\vspace{-15pt}
\end{figure}

\fref{fig:zoi_av_interv} shows the box-plots, with interquartile range (IQR) and population median (in red), of the success ratio obtained by simulating the strategies from the DeepFloat approach. Outliers exit the $Q_x\pm 1.5\;IQR$, with $x\in\{1,3\}$ quartile $Q$. As shown, our strategies allow the system to achieve in both intervals the minimum value of success ratio ($0.9$).  

As we have observed, strategies emerging from our DeepFloat approach try to minimize, for each link, the difference between the amount of seeding required at the beginning of each interval, and the mean availability at the previous interval. 
It produces strategies that tune the distribution of content in an interval taking into account the seeding requirements at the following interval. 
%
 With the given choice of coefficients ($~\beta,\delta=1$), 
 in the whole set of simulations 
 no seeding support has been necessary at the beginning of the second interval, in any link. This suggests the effectiveness of our approach in accurately controlling content availability in space and time as a function of resource cost, and in particular, in  
 decreasing infrastructure support whenever its cost becomes higher than the cost of using hosts resources.

\begin{figure}[t!]
  \centering
\begin{subfigure}[]{\includegraphics[width=0.21\textwidth]{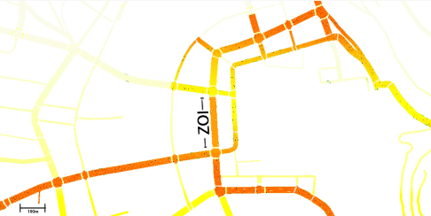}\label{fig:t1_map_replic}}
\end{subfigure}
\begin{subfigure}[]{\includegraphics[width=0.21\textwidth]{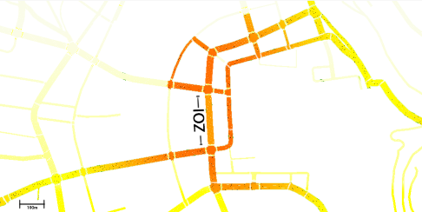}\label{fig:t2_map_replic}}
\end{subfigure}
\begin{subfigure}{\includegraphics[width=0.03\textwidth]{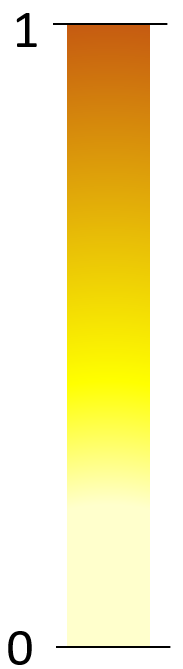}}
\end{subfigure}
\caption{Luxembourg city center replication strategy at t1 = (7 AM - 7:30 AM) (a) and t2 = (7:30 AM - 8 AM) (b).}\label{fig:map_replic}
\end{figure}

\begin{figure}[t!]
  \centering
\begin{subfigure}[]{\includegraphics[width=0.21\textwidth]{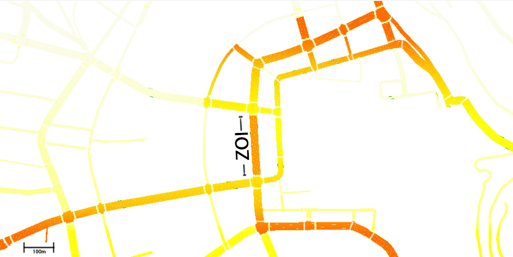}\label{fig:t1_map_storage}}
\end{subfigure}
\begin{subfigure}[]{\includegraphics[width=0.21\textwidth]{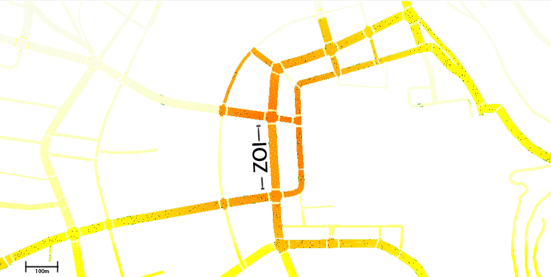}\label{fig:t2_map_storage}}
\end{subfigure}
\begin{subfigure}{\includegraphics[width=0.03\textwidth]{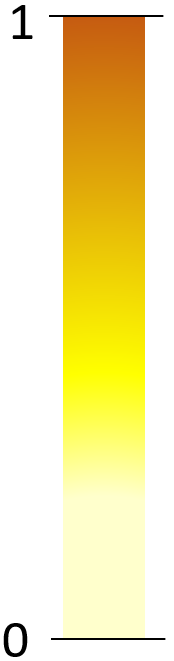}}
\end{subfigure}
\caption{Luxembourg city center storage strategy t1 = (7 AM - 7:30 AM) (a) and t2 = (7:30 AM - 8 AM) (b).}\label{fig:map_storage}
\vspace{-10pt}
\end{figure}

Another evidence of the dynamic adaptation of the temporal strategy to traffic conditions is shown in Figure~\ref{fig:map_replic} and~\ref{fig:map_storage}. To handle the increase in vehicle density while meeting the performance target at $t2$, the system effectively adapts both replication and caching areas. Figures~\ref{fig:t2_map_replic} and~\ref{fig:t2_map_storage} show how the system activates only those resources near the ZOI that are necessary to achieve the performance target. On the other hand, the system lowers replication and caching probabilities where possible, to avoid unnecessary resource usage (see the north-west areas in Figures~\ref{fig:map_replic} and~\ref{fig:map_storage}).

\subsection{Time considerations}
 In the Luxembourg scenario, characterized by $167$ links, and on a I7 desktop PC with 16 GB of RAM, the first two phases of our approach (training set buildup and CNN training) together took $3$s for a training set of $10^3$ batches, and about $3$ min for $10^6$ batches. Hence, at least in the considered scenario (i.e., a whole city center), and with the given choices of links and parameters, the computational load of the offline phase does not require a large number of resources. 
As for the bootstrap phase, in all our experiments, the computation of the FC strategy by the CNN always took less than one second, suggesting that, despite the large road grid and the large size of the training set considered, our approach is fit for real-time operation while keeping an excellent level of accuracy.



\section{Conclusions}
\label{concl}
In this work, we have outlined a Deep Learning approach for efficient FC dimensioning, which exploits a Convolutional Neural Network to modulate over time the parameters governing an FC service. We have shown on realistic, measurement-based scenarios that our approach efficiently shapes the communication area and optimizes replication, caching, and seeding strategies while achieving the desired QoS targets over time. Among the possible extensions to our approach, we intend to introduce temporal components in the learning models, and to consider those to scenarios in which mobility patterns are influenced by the spreading of content (e.g., traffic jams notification).


\section{Acknowledgment}
This research was supported/partially supported by the Swiss National Science Foundation (SNSF, project CONTACT, no. 164205), by Hasler MOBNET, and by COST RECODIS.

\bibliographystyle{IEEEtran}
\bibliography{IEEEabrv,main}

\end{document}